\documentclass[preprint,prd,aps,showpacs,preprintnumbers,amsmath,amssymb]{revtex4-1}
\usepackage{graphics,epsfig,subfigure}
\usepackage{diagbox}
\usepackage[usenames]{color}
\usepackage{color}
\usepackage{graphicx}
\usepackage{amsfonts}
\usepackage{indentfirst}
\usepackage{booktabs}
\usepackage{hyperref}
\hypersetup{hypertex=ture,backref=true,colorlinks=ture,linkcolor=blue,anchorcolor=blue,citecolor=blue}
\usepackage{float}
\usepackage{multirow}
\usepackage[section]{placeins}

\begin{document}

\title{\Large \bf Ellis wormhole with nonlinear electromagnetic field}
\author{Xin Su}
\author{Chen-Hao Hao}
\author{Yong-Qiang Wang\footnote{yqwang@lzu.edu.cn, corresponding author}}
\affiliation{ $^{1}$Lanzhou Center for Theoretical Physics, Key Laboratory of Theoretical Physics of Gansu Province,
	School of Physical Science and Technology, Lanzhou University, Lanzhou 730000, China\\
	$^{2}$Institute of Theoretical Physics $\&$ Research Center of Gravitation, Lanzhou University, Lanzhou 730000, China}

\begin{abstract}
In this paper, we present the spherically symmetric wormhole in Einstein’s gravity coupling phantom field and nonlinear electromagnetic field. Numerical results show that this solution violates the Null Energy Condition (NEC), and as the parameters change, the ADM mass of the entire spacetime changes from positive to negative. In addition, we analyze the light ring (LR) of the solution and demonstrate the astronomical observation properties. Especially when negative mass appears, the general LR will not appear, only a ``special unstable LR" exists at the throat, which is caused by the repulsive effect of the negative mass on both sides of the wormhole. Finally, we draw the embedding diagram to reflect the geometric characteristics of the wormhole. 
\end{abstract}

\maketitle

\section{INTRODUCTION}\label{sec1}

In general relativity, singularities are unavoidable in most physically relevant black hole solutions. But since matter cannot be infinitely compressed, singularities is considered unphysical. Thus, much effort has been ongoing to resolve the singularity problem. In the early 20th century,  J. M. Bardeen proposed a regular black hole model without singularities \cite{Lan:2023cvz}. Subsequently, S. A. Hayward introduced another regular black hole model while investigating black hole formation and evaporation \cite{Hayward:2005gi}. Apart from the regular black hole, another spacetime structure in GR that could eliminate singularities is the traversable wormhole. A wormhole is a theoretically predicted structure that connects two independent points in spacetime, potentially enabling interstellar travel. The concept of wormholes dates back to 1916 when German physicist L. Flamm first proposed it. In 1935, A. Einstein and N. Rosen extended the idea by introducing the Einstein-Rosen bridge, which represents a non-traversable wormhole \cite{Einstein:1935tc}. After that, the structure of the traversable wormhole solution attracted attention \cite{Ellis:1973yv,Bronnikov:1973fh}. Especially by 1988, the outstanding work of M. Morris and K. Thorne enabled the traversable wormhole solution to be fully explained \cite{Morris:1988cz}. They emphasized that to construct a traversable wormhole, it is necessary to introduce exotic matter with negative energy density, which can be achieved by introducing a phantom field with a negative kinetic energy term. In recent years, wormholes coupling other matter fields have attracted more and more attention\cite{Dzhunushaliev:2013lna,Charalampidis:2013ixa,Hoffmann:2017vkf,Dzhunushaliev:2014bya,Aringazin:2014rva,Hoffmann:2017jfs,Ding:2023syj,Su:2023xxk,Hao:2023igi}, and some novel phenomena have emerged.

To make the wormhole traversable, it seems necessary to violate NEC. This violation of energy conditions originally came from the introduction of exotic matter. Alternatively, by limiting exotic matter to the throat, a thin shell wormhole could be created with its reduced use \cite{Visser:1989kh,Poisson:1995sv,Eiroa:2003wp,Lobo:2004rp,Gravanis:2007ei}. Subsequent research has shown that modified gravity or Einstein Maxwell Dirac theory can be used to construct traversable wormholes to avoid introducing phantom exotic matter \cite{Lobo:2009ip,Kanti:2011jz,Rahaman:2013qza,Bronnikov:2015pha,Shaikh:2016dpl,Rani:2016xpa,Sahoo:2017ual,Dai:2020rnc,DeFalco:2021ksd,DeFalco:2023twb,Blazquez-Salcedo:2020czn,Konoplya:2021hsm,Wang:2022aze}. Energy conditions initially seemed sacred and inviolable, representing all physical reasonableness. By setting energy conditions,  it is possible to prove the singularity theorem (dependent on SEC), the topological censorship theorem (dependent on NEC), etc \cite{Hawking:1970zqf,Witten:1981mf,Hawking:1991nk}. Over years, Perspectives on energy conditions have evolved. More and more theories and models have broken the energy conditions \cite{Barcelo:2002bv}. Violation of energy conditions will lead to the emergence of many novel physical phenomena, such as the emergence of negative mass \cite{Bonnor89} and the curvature engine caused by the distortion of space and time \cite{Alcubierre:1994tu}. In fact, Some propose that the manifestation of negative mass matter is reasonable. For example, negative mass as a candidate for dark energy, can be used to explain the accelerated expansion of the universe \cite{Farnes:2017gbf}. \cite{Lu:2015cqa} constructed the first vacuum quadratic gravity non-Schwarzschild black hole solution through numerical calculations and obtained a negative mass black hole within a specific parameter range. In addition, rotating wormhole solutions supported by a complex phantom scalar fiel is considered in \cite{Chew:2019lsa}, and a negative mass wormhole was obtained. Rencently, \cite{Hao:2023kvf} has coupled the nonlinear electromagnetic field under the Bardeen model with wormhole spacetime, and the results show that within a specific parameter range, the entire system will exhibit negative mass. There are some other works related to negative mass\cite{Goldstein:2017rxn,Huang:2022urr,Mbarek:2014ppa,Dzhunushaliev:2024iag}.

From an observational perspective, the  Event Horizon Telescope (EHT) observed optical images of the supermassive black hole located at the center of the solar system, which is a major achievement in the field of astronomy \cite{EventHorizonTelescope:2019dse,EventHorizonTelescope:2019uob,EventHorizonTelescope:2019jan,EventHorizonTelescope:2019ths,EventHorizonTelescope:2019pgp}. For  a black hole, the event horizon that defines its characteristics is not visible, but the light ring (LR), which represents a closed circular orbit of photons, is an important observable characteristic of a black hole \cite{Cardoso:2021sip,Rosa:2023hfm,Cunha:2020azh}. But this is not unique to black holes. Some other compact objects, such as Boson stars, Proca stars \cite{Rosa:2022tfv,Rosa:2022toh}and gravity stars can all have light rings, of course wormholes. For symmetrical wormholes, there will always be an unstable LR at the throat \cite{Xavier:2024iwr}. For asymmetrical wormholes, the LR may not necessarily exist at the throat. There can be a LR on one side of the universe but not on the other side \cite{Huang:2023yqd}. If a negative mass wormhole really exists, would it have LR like a positive mass wormhole? This is a question worth exploring. In our model, we coupled the wormhole model with the nonlinear electromagnetic field in Hayward spacetime, and the results showed that negative ADM mass appears within a certain parameter range. In addition, we focused on the distribution of the LR and we found that especially when negative mass appears, the distribution of the light rings will be different from positive mass case.

This paper is structured as follows. In Sec.~\ref{sec2}, we construct a traversable wormhole model of a phantom scalar field
and a nonlinear electromagnetic field coupled with gravity. In Sec.~\ref{sec3}, we mainly give the boundary conditions required for the solution and the quantity of interest. We mainly classify and summarize our numerical results in Sec.~\ref{sec4}. The last section is a summary and outlook of further research.\raggedbottom

\section{MODEL}\label{sec2}

We consider Einstein’s gravity minimally coupled phantom scalar field and the nonlinear electromagnetic field of the Hayward model, the action is given by
\begin{equation}
S=\int \sqrt{-g} d^{4} x\left(\frac{R}{2 \kappa}+\mathcal{L}_{H}+\mathcal{L}_{P}\right),
\end{equation}
the term $\mathcal{L}_{H}$ and $\mathcal{L}_{P}$ are the Lagrangians defined by
\begin{equation}
\mathcal{L}_{H}=\frac{3}{2s} \frac{\left(\sqrt{2 {q}^{2} \mathcal{F}}\right)^{3}}{\left(1+\left(\sqrt{2 {q}^{2}\mathcal{F}}\right)^{3 / 2}\right)^{2}} ,
\end{equation}
\begin{equation}
\mathcal{L}_{2}=\nabla_{a} \Phi \nabla^{a} \Phi,
\end{equation}
where $\kappa$ represents the coupling constant, $R$ is the Riccci scalar, $\mathcal{F}=\frac{1}{4} F_{\mu \nu} F^{\mu \nu}$ with the elextromagnetic field $F_{\mu \nu}=\partial_{\mu} A_{\nu}-\partial_{\nu} A_{\mu}$, in which $A_{\mu}$ is the electromagnetic 4-potential. Here, $\Phi$ denotes the phantom scalar field. The constants $q$ and $s$ are two independent parameters, where $q$ represents the magnetic charge.

Varying the action (1) with respect to the metric, electromagnetic field, and phantom field, we derive the following equations
\begin{equation}
R_{\mu \nu}-\frac{1}{2} g_{\mu \nu} R=\kappa (T_{\mu \nu}^{H}+T_{\mu \nu}^{P}),
\end{equation}
\begin{equation}
\nabla_{a}\left(\frac{\partial \mathcal{L}^{H}}{\partial \mathcal{F}} F^{a b}\right)=0,
\end{equation}
\begin{equation}
\square \Phi=0,
\end{equation}
here, $R_{\mu \nu}$ is the Ricci curvature tensor, $T_{\mu \nu}^{H}$ and $T_{\mu \nu}^{P}$
represent the energy-momentum tensors for the electromagnetic field and the phantom field,  respectively
\begin{equation}
T_{\mu \nu}^{H}=-\frac{\partial \mathcal{L}_{H}}{\partial \mathcal{F}} F_{\mu \sigma} F_{\nu}{ }^{\sigma}+g_{\mu \nu} \mathcal{L}_{H},
\end{equation}
\begin{equation}
T_{\mu\nu}^{(P)}=\partial_{\mu} \Phi^{*} \partial_{\nu} \Phi+\partial_{\nu} \Phi^{*} \partial_{\mu} \Phi-g_{\mu\nu}\left[\frac{1}{2} g^{\mu\nu}\left(\partial_{\mu} \Phi^{*} \partial_{\nu} \Phi+\partial_{\nu} \Phi^{*} \partial_{\mu} \Phi\right)\right] .
\end{equation}

In this paper, we consider general static spherically symmetric sulution with a wormhole, and adopt the ansatz as follows \cite{Dzhunushaliev:2014bya}
\begin{equation}
d s^{2}=-e^{B} d t^{2}+C e^{-B}\left[d r^{2}+h\left(d \theta^{2}+\sin ^{2} \theta d \varphi^{2}\right)\right],
\end{equation}
where $B$ and $C$ are both functions of the radial coordinate $r$, $h=r^{2}+r_{0}^{2}$, $r_{0}$ is the throat size. The range of $r$ is from negative infinity to positive infinity. When $r$ approaches positive and negative infinity, it corresponds to two asymptotically flat spacetimes.  In addition, we use the following ansatzes of the electromagnetic field and the phantom field
\begin{equation}
A=q \cos (\theta) d \varphi,
\end{equation}
\begin{equation}
\Phi=\phi(r).
\end{equation}

Substituting the above ansatzes (9) - (11) into the field equations (4)and (6), we can get the following equations

\begin{equation}
\frac{6 e^{2 B}\kappa q^{6} c^{2}\left(e^{\frac{7}{4}} q^{3}-2\left(h^{2} c\right)^{3 / 2}\right)}{s\left(e^{\frac{3 B}{2}} q^{3}+\left(h^{2} c\right)^{3 / 2}\right)^{3}}+\frac{1}{2} B^{\prime} C^{\prime}+C\left(\frac{2 r B^{\prime}}{h^{2}}+B^{\prime \prime}\right)=0,
\end{equation}

\begin{equation}
\frac{6 e^{2 B}\kappa q^{6} c^{2}\left(2 e^{\frac{7 B}{4}} q^{3}-\left(h^{2} c\right)^{3 / 2}\right)}{s\left(e^{\frac{3 B}{2}} q^{3}+\left(h^{2} c\right)^{3 / 2}\right)^{3}}+\frac{3 r C^{\prime}}{h^{2}}-\frac{C^{\prime 2}}{2 C}+C^{\prime \prime}=0,
\end{equation}

\begin{equation}
\left(h \sqrt{C} \phi^{\prime}\right)^{\prime}=0.
\end{equation}
Integrating the equation (14) leads to the following equation
\begin{equation}
\phi^{\prime}=\frac{\sqrt{\mathcal{D}}}{h \sqrt{C}}.
\end{equation}
The constant $D$ is the scalar charge of the phantom field. Substituting equation (15) into Einstein's equation component can eliminate the $\phi^{'}$term of the phantom field, and obtain the specific expression of $D$ as follows
\begin{equation}
\mathcal{D}=\frac{h^{4} c}{\kappa}\left(\frac{r_{0}^{2}}{h^{4}}-\frac{3 e^{2 B} q^{6} c}{s\left(e^{\frac{3 B}{2}} q^{3}+\left(h^{2} c\right)^{3 / 2}\right)^{2}}+\frac{B^{\prime 2}}{4}-\frac{r C^{\prime}}{h^{2} c}-\frac{c^{\prime 2}}{4 c^{2}}\right).
\end{equation}
Solving the OED equations (12) and (13) numerically, we can get all information about metric functions $B$, $C$, and the scalar charge $D$ of the phantom field.

\section{BOUNDARY CONDITIONS AND QUANTITIES OF INTEREST}\label{sec3}
To solve the coupled non-linear equations (12), (13) in asymptotically flat spacetime, we give appropriate boundary conditions for the functions at infinity as follows
\begin{equation}
B=0(r\to \infty ),\; C(r\to \infty )=1,\; \partial_{r} C(r\to \infty )=0 .
\end{equation}
However, in our model, there are two field functions whose symmetry is not consistent, so we no longer apply additional boundary conditions at $r = 0$.

The ADM mass $M$ is the key quantity we are interested in, which is encoded in the asymptotic expansion of
metric components
\begin{equation}
g_{t t}=-1+\frac{2 M}{r}+\cdots.
\end{equation}

Due to the symmetry of two independent asymptotically flat universes about the wormhole throat, the ADM mass can also be obtained from the Komar expression.
\begin{equation}
M=\int_{\Sigma} R_{\mu \nu} n^{\mu} \xi^{\nu} d V.
\end{equation}

In this paper, the ADM mass obtained by the above two methods is the same.

\section{NUMERICAL RESULTS}\label{sec4}
Here we present our numerical results, with $s$ fixed to 1. For other $s$ values, there are consistent conclusions. To facilitate calculation,  we compactify the coordinates by introducing a new radial coordinate $x$
\begin{equation}
r=\tan(\frac{\pi}{2}x) ,
\end{equation}
$r \in(-\infty,+\infty)$ corresponding to $x \in(-1,+1)$.

\begin{figure}
  \begin{center}
\subfigure{\includegraphics[width=0.48\textwidth]{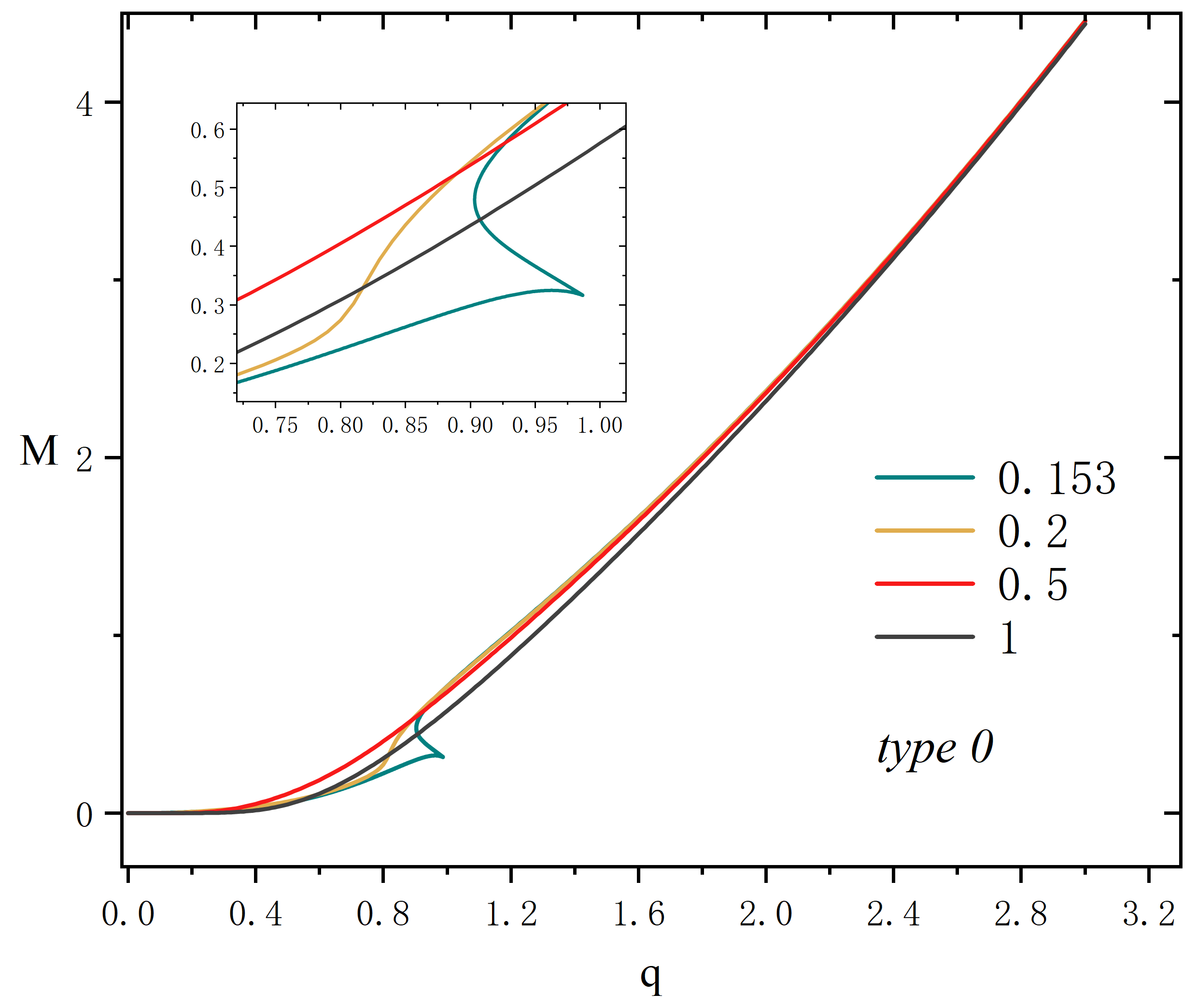}}
\subfigure{\includegraphics[width=0.5\textwidth]{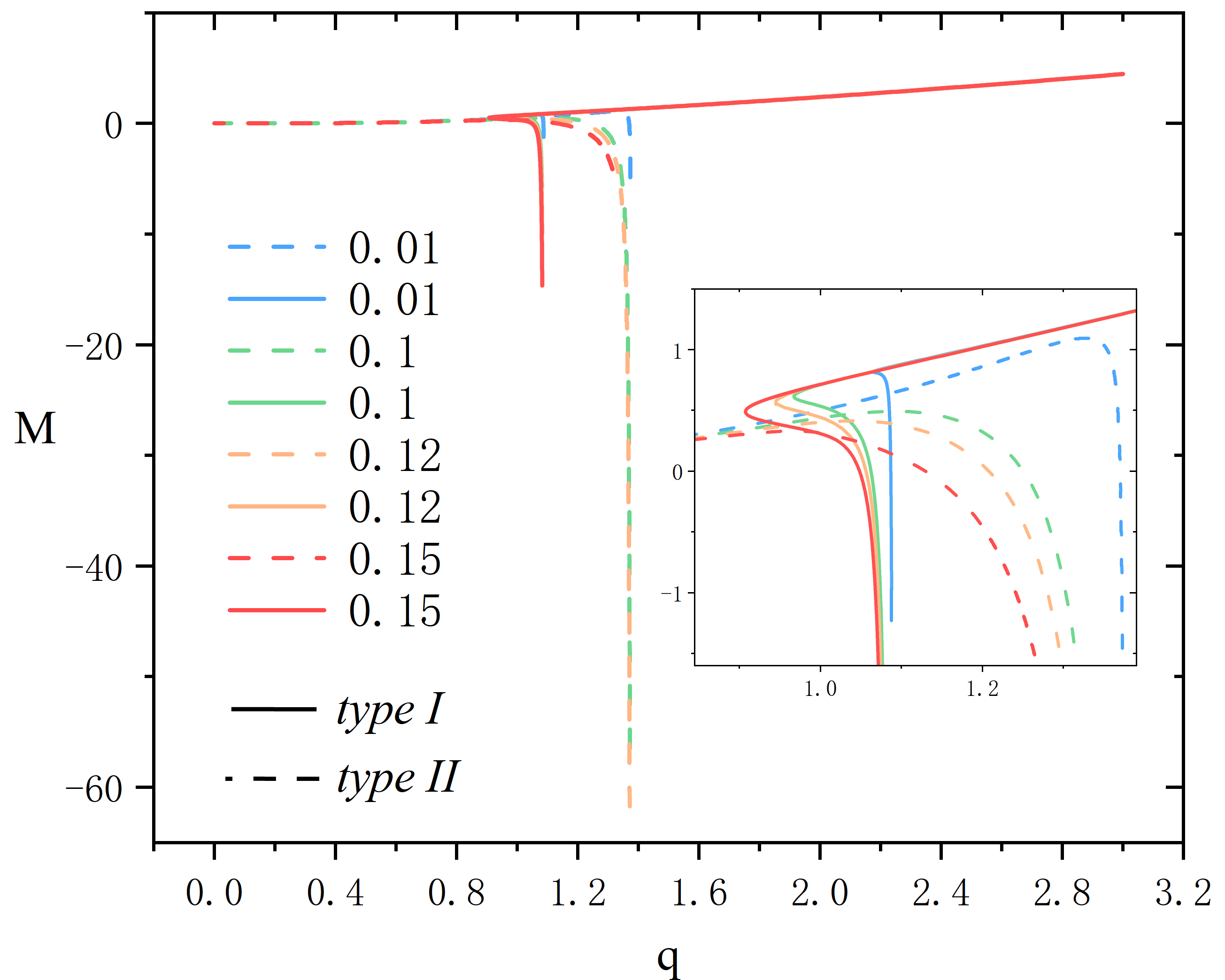}}
  \end{center}
\caption{The mass $M$ as functions of $q$ for different values of $r_{0}$. The left figure corresponds to the \textit{type 0} solution where the ADM mass is always positive. The right figure corresponds to two types of solutions which have negative mass region, the solid line represents \textit{type I}, and the dotted line represents \textit{type II}.}
\label{mm}
\end{figure}

\subsection{ADM mass $M$, scalar charge $D$ and energy condition}

By varying throat size $r_{0}$, we obtain a series of relationships between the corresponding ADM mass and $q$ as shown in Fig. \ref{mm}. Due to the symmetry of the wormhole, the ADM mass on both sides of the throat is identical. Thus, only the ADM mass of one side is shown. For large $r_{0}$, the ADM mass is always positive as shown in the left panel of Fig. \ref{mm}, marked as \textit{type 0}. When $q=0$, the electromagnetic field vanishes and the ADM mass is zero corresponding to an Ellis wormhole. As $q$ increases, the ADM mass increases monotonically. When $r_{0}$ decreases to about 0.2, the curve remains monotonic, but within a range where $q$ takes a smaller value, the curve exhibits a ``fluctuation". The smaller $r_{0}$ is, the more obvious the curve fluctuation is. At $r_{0}=0.153$, the $M$ curve fluctuation is the most significant. When $r_{0}$ is less than 0.153, a crucial feature emerges: the $M$ curve will split into two different solutions marked as \textit{type I} and \textit{type II} in the right panel of Fig. \ref{mm}. In the \textit{type I} solution, $M$ initially decreases with decreasing $q$, then sharply turns and within a very narrow range, it becomes negative as $q$ increases. In the \textit{type II} solution, $M$ initially rises as $q$ increases from 0, however, once it reaches a specific value, $M$ sharply declines with further increases in $q$, eventually becoming strongly negative. Additionally, we find that as $r_{0}$ decreases, the absolute value of the negative mass that the two solutions can finally achieve first increases and then decreases. If $r_{0}$ is smaller, neither solution can give the negative ADM mass.

\begin{figure}
  \begin{center}
\subfigure{\includegraphics[width=0.48\textwidth]{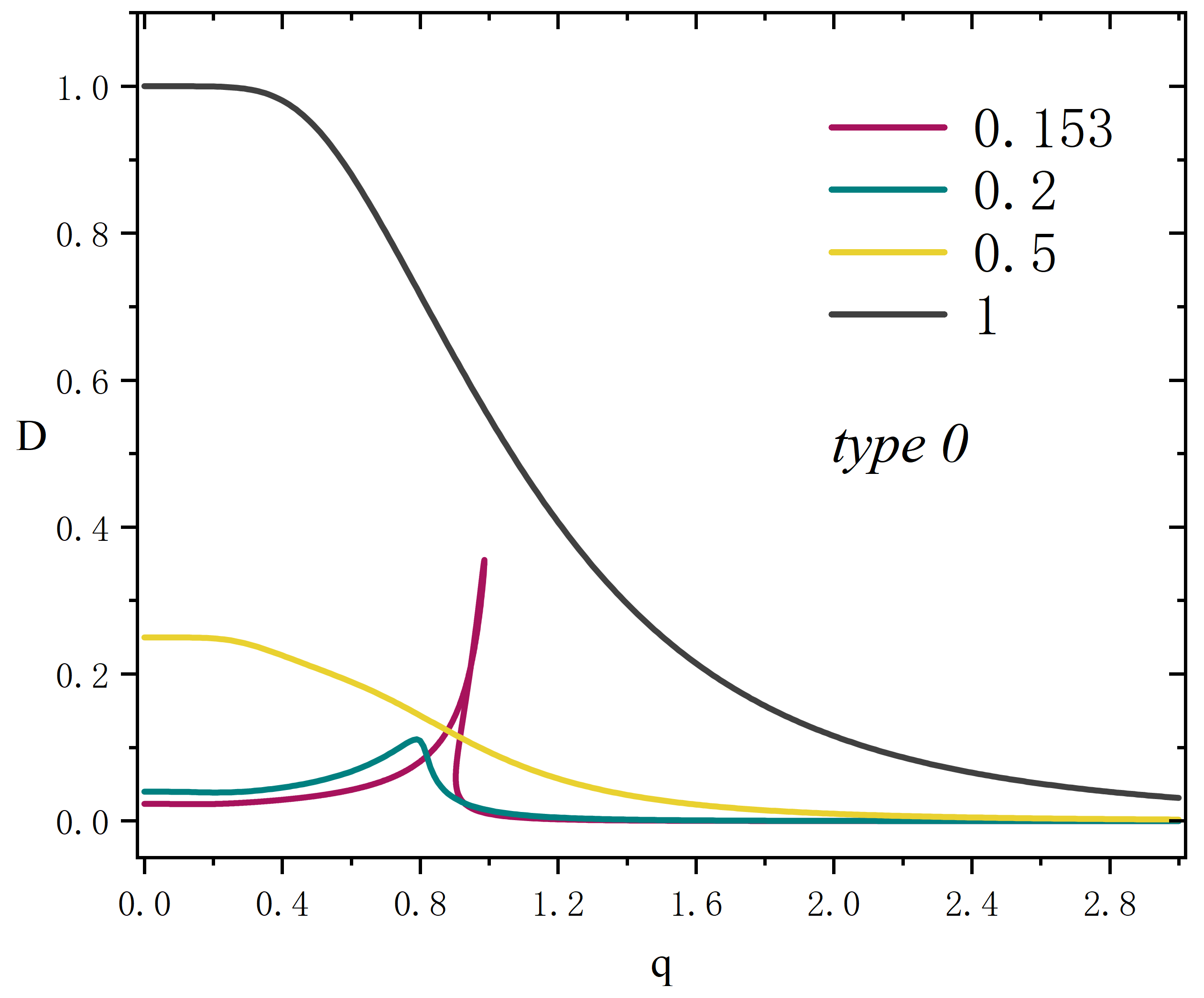}}
\subfigure{\includegraphics[width=0.5\textwidth]{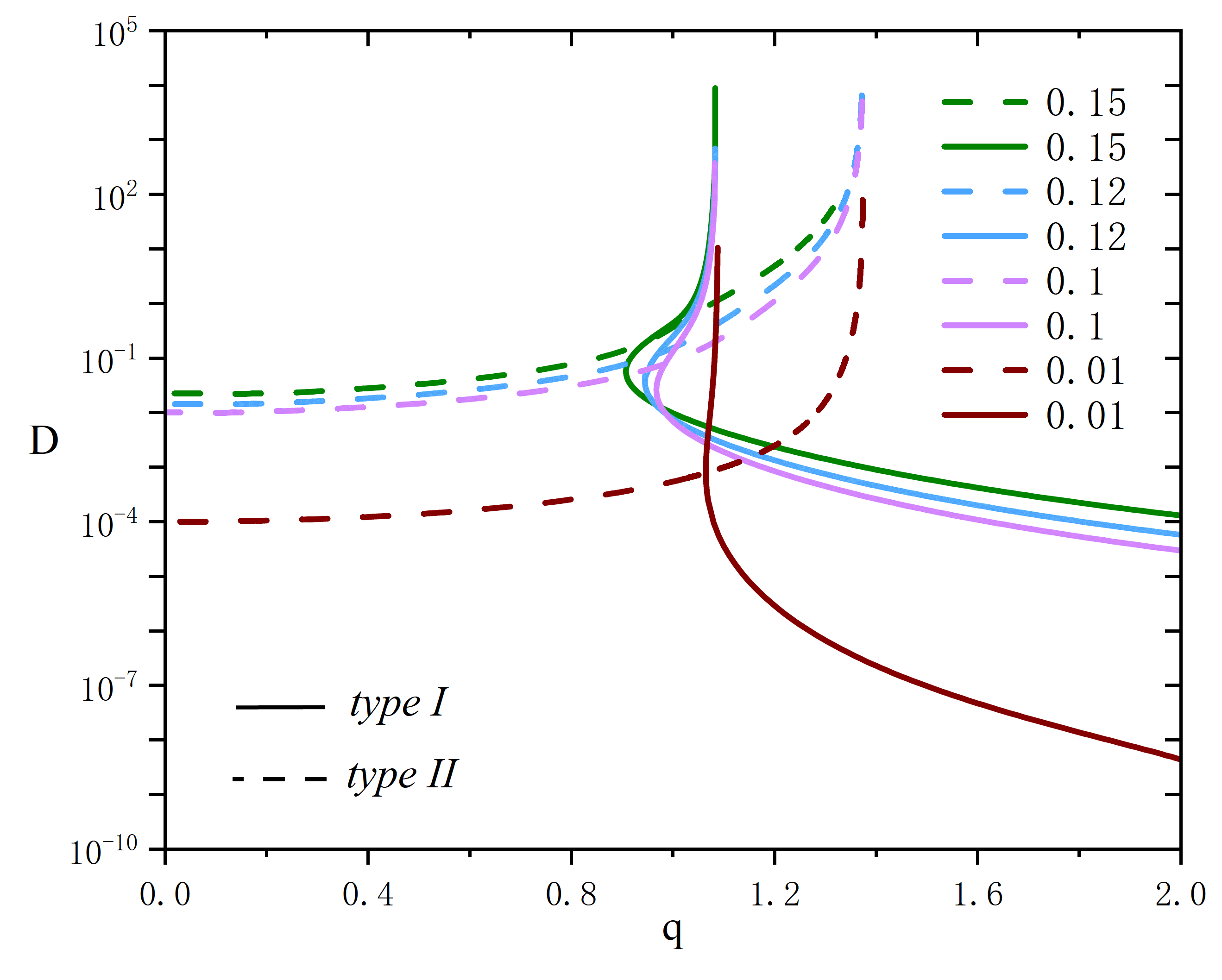}}
  \end{center}
\caption{The scalar charge $D$ as functions of $q$ for different values of $r_{0}$. The left figure corresponds to the \textit{type 0} solution, the solid lines in the right figure correspond to the \textit{type I} solution, and the dashed line corresponds to the \textit{type II} solution.}
\label{dd}
\end{figure}

In Fig. \ref{dd}, we show the relationship between the scalar charge $D$ and the $q$ for different values of the throat size $r_{0}$. On the one hand, the scalar charge $D$ can be used to verify the accuracy of numerical calculations. Specifically, we calculate the value of $D$ at different $x$ positions and observe the fit of the function curves. We only retain the results with a difference of less than $10^{-4}$. On the other hand, the value of $D$ can reflect the amount of phantom matter. From the graphs, we can see that as $r_{0}$ increases, the value of $D$ becomes larger. Especially, in the area where negative mass appears in the solution the magnitude of the $D$ is extremely large, indicating that phantom matter is highly concentrated. The highly concentrated phantom field may cause the Komar integration of ADM mass to be negative.

\begin{figure}
  \begin{center}
\subfigure{\includegraphics[width=0.48\textwidth]{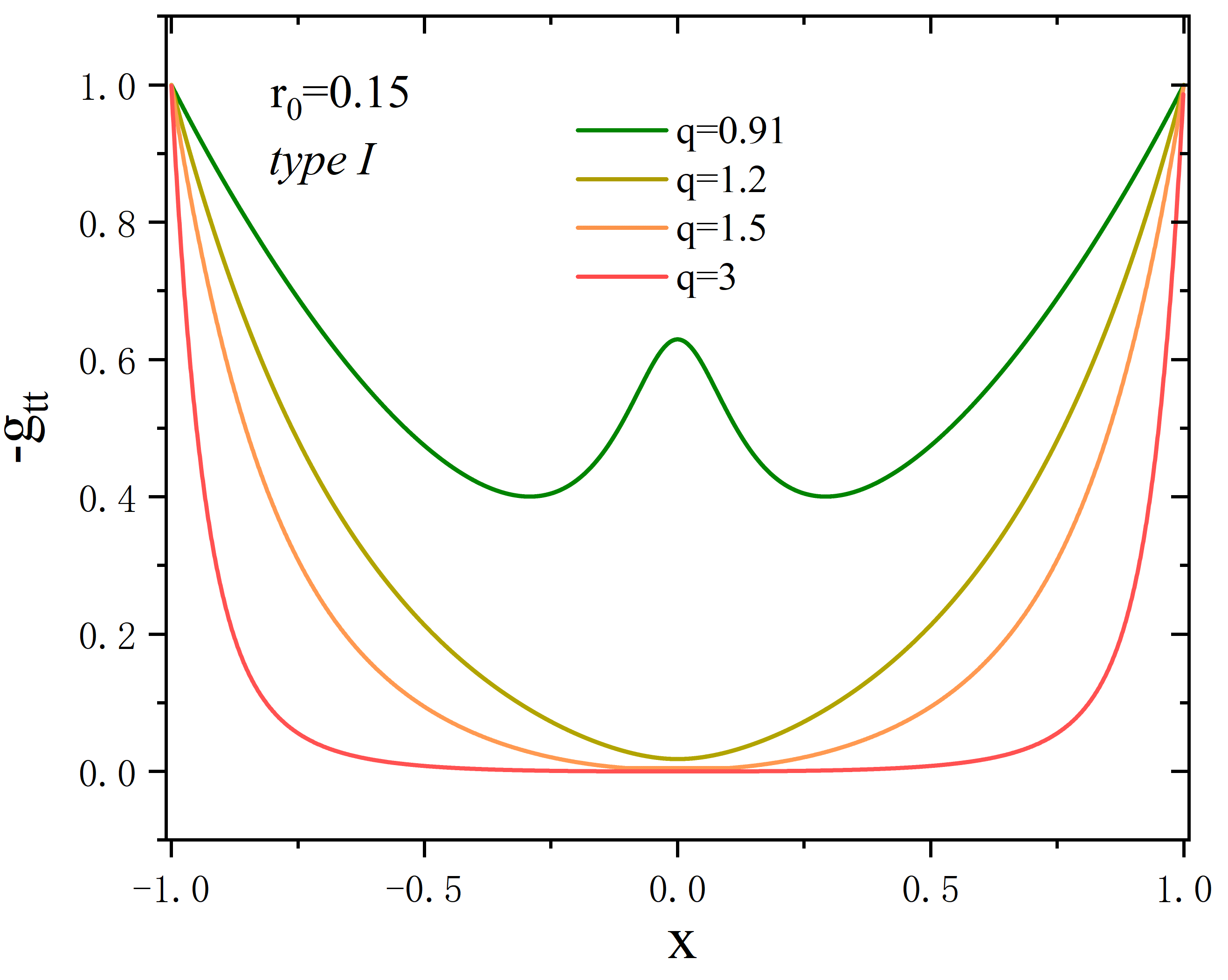}}
\subfigure{\includegraphics[width=0.48\textwidth]{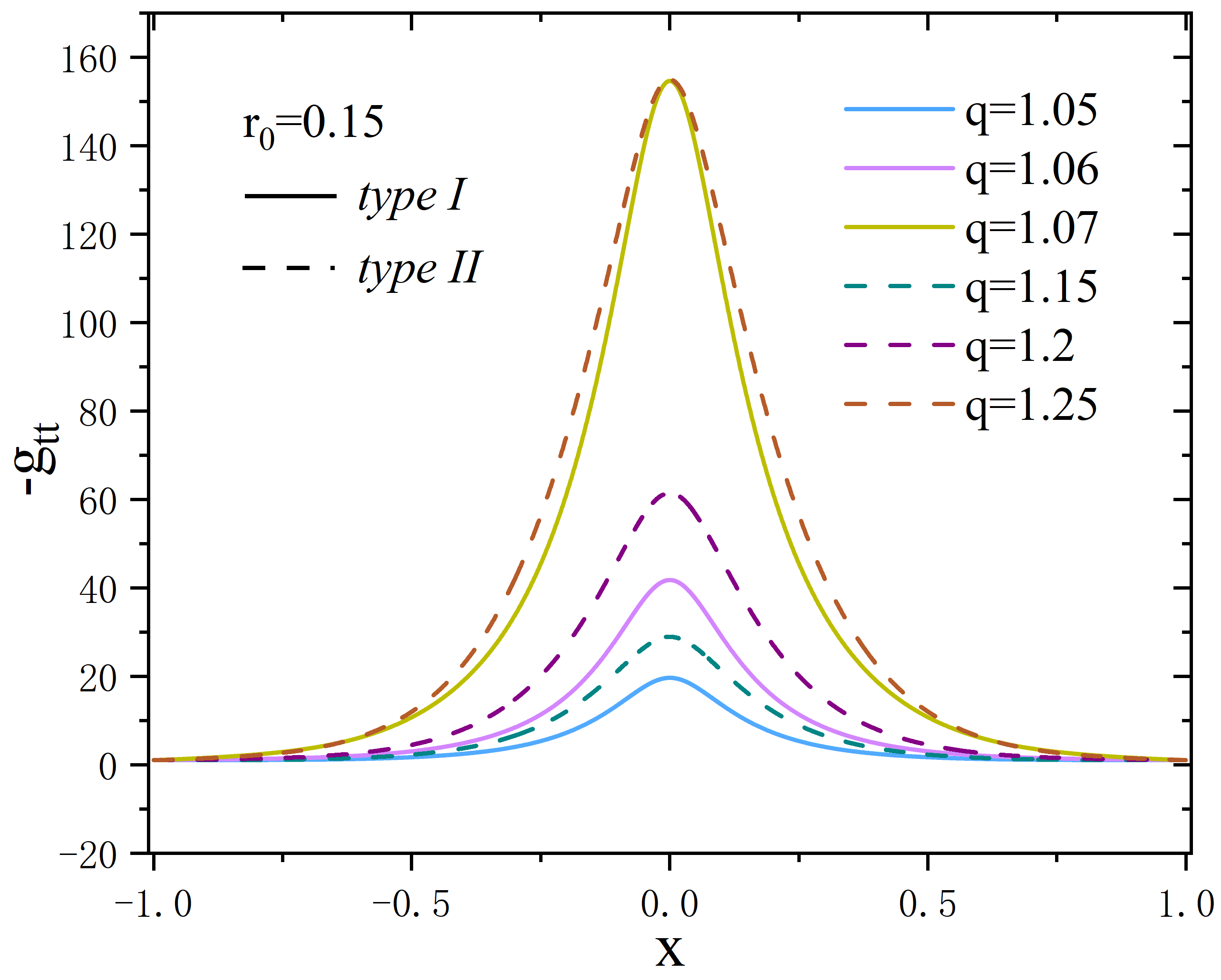}}
  \end{center}
\caption{ The functional relationship of $-g_{tt}$ concerning $x$, with the parameter $s$ fixed at 1, $r_{0}$ fixed at 0.15 and $q$ with some different values. The solid line represents the \textit{type I} solution, and the dashed line represents the \textit{type II} solution.}
\label{gtt}
\end{figure}

Due to the emergence of negative mass, the metric function will reflect some unusual spacetime characteristics. In Fig. \ref{gtt} we calculate the metric component $g_{tt}$ with $r_{0}$ fixed to 0.15. The left panel corresponds to the positive mass part of the \textit{type I} solution. When $q$ takes a larger value, $-g_{tt}$ reaches the minimum value at $x=0$, and the minimum value is very close to zero which means a horizon is about to appear. As $q$ decreases near the region of negative mass, the value of $-g_{tt}$ increases and the minimum value is not at $x=0$. The right panel shows the $-g_{tt}$ curves of the negative mass region for \textit{type I} and \textit{type II}. As the negative mass magnitude increases, $-g_{tt}$ further increases, particularly at $x=0$, the throat of the wormhole, where all $-g_{tt}$ values exceed 1. At this time, observers at infinity will feel an obvious blue shift effect.

\begin{figure}
  \begin{center}
\subfigure[~]{\label{n1}
\includegraphics[width=0.32\textwidth]{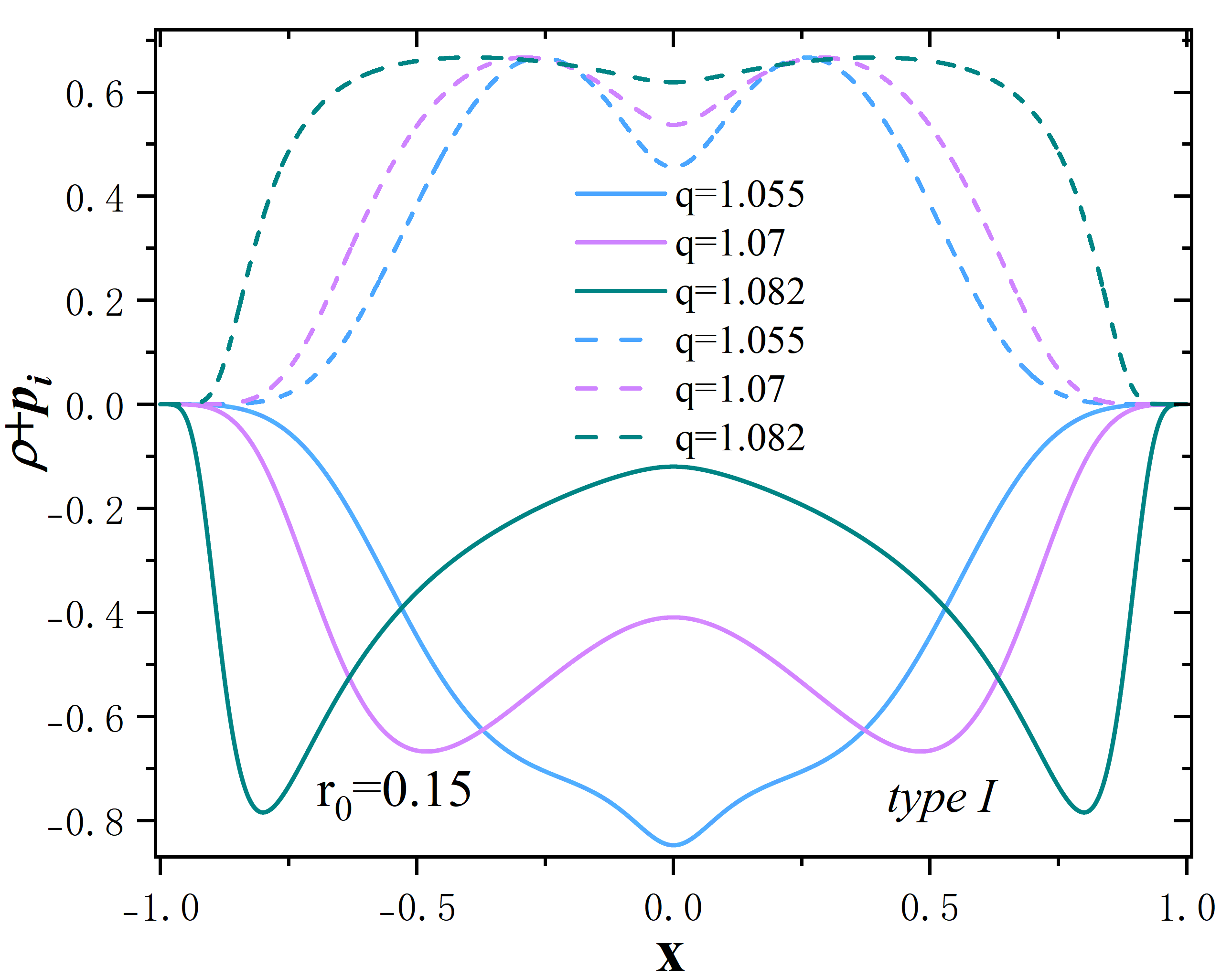}}
\subfigure[~]{\label{n2}
\includegraphics[width=0.32\textwidth]{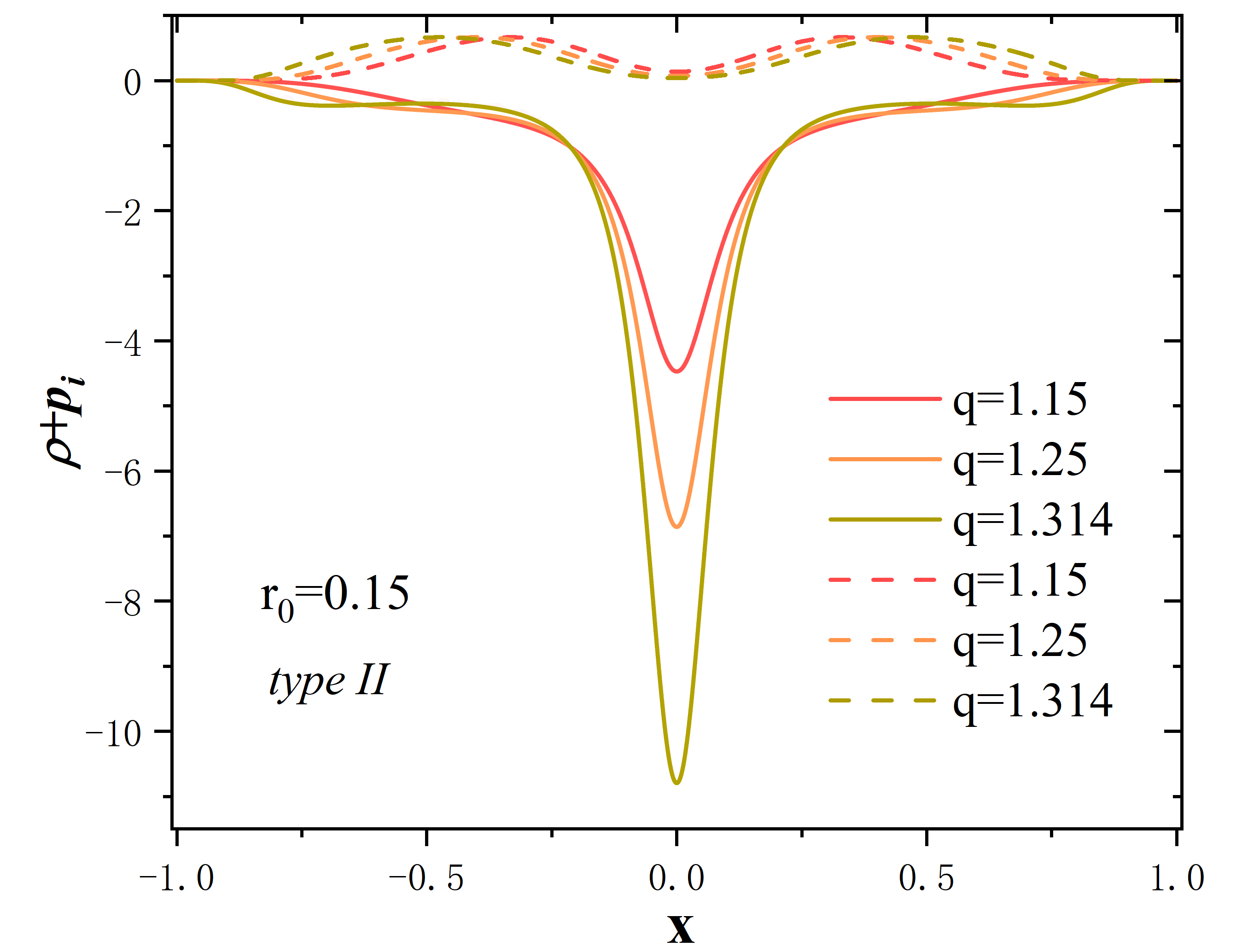}}
\subfigure[~]{\label{n3}
\includegraphics[width=0.3\textwidth]{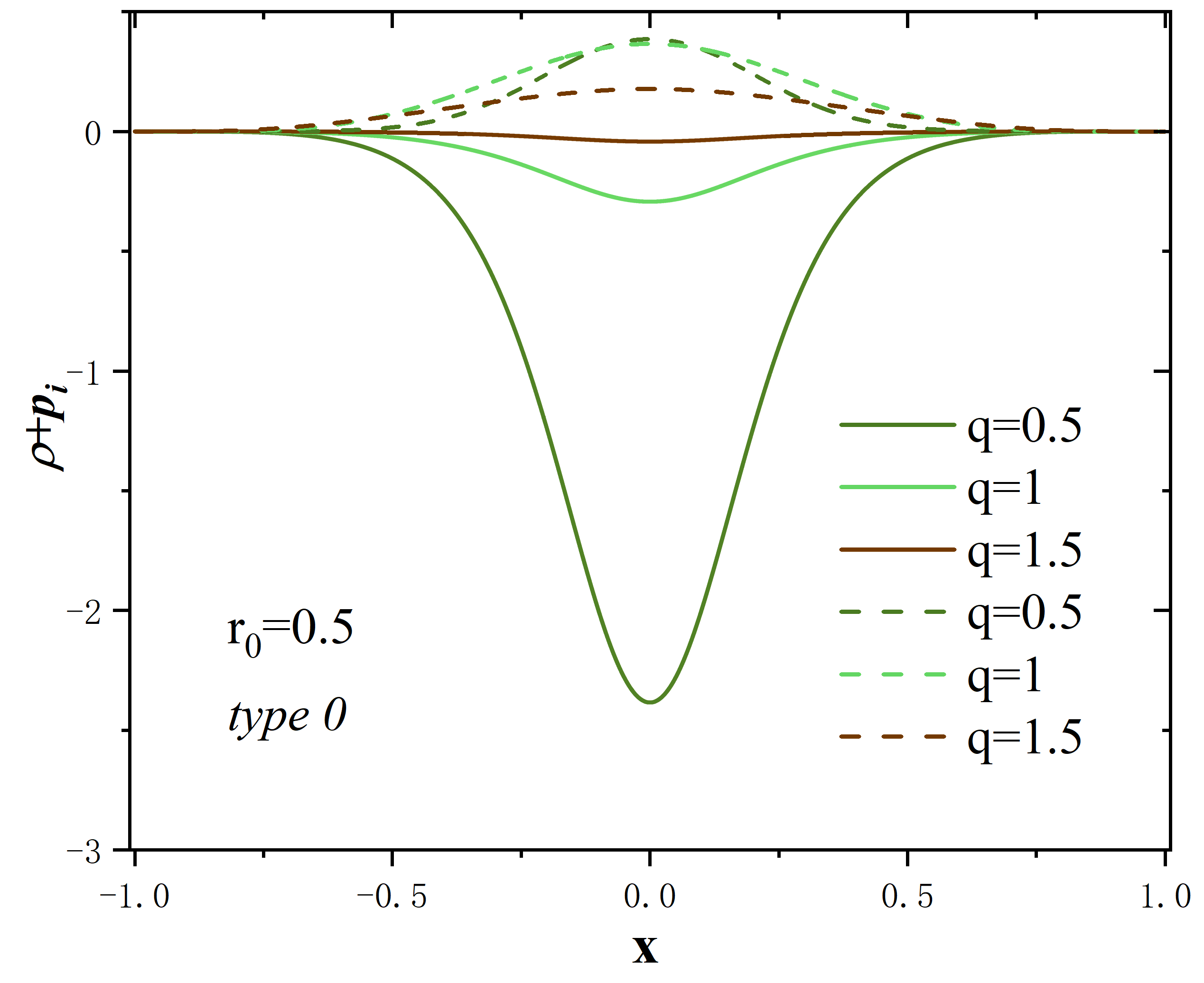}}
  \end{center}
\caption{ NEC versus coordinate $x$ for different $q$ . Fig. (a) and (b) correspond to the negative mass regions of \textit{type I} and \textit{type II} solutions when $r_{0}$ = 0.15. Fig. (c) represents the positive mass solution \textit{type 0} for $r_{0}$ = 0.5. The solid line denotes $\rho + p_{1}$, the dotted line signifies $\rho + p_{2}$. }
\label{nec}
\end{figure}

\begin{figure}
  \begin{center}
\subfigure[~]{\label{r1}
\includegraphics[width=0.33\textwidth]{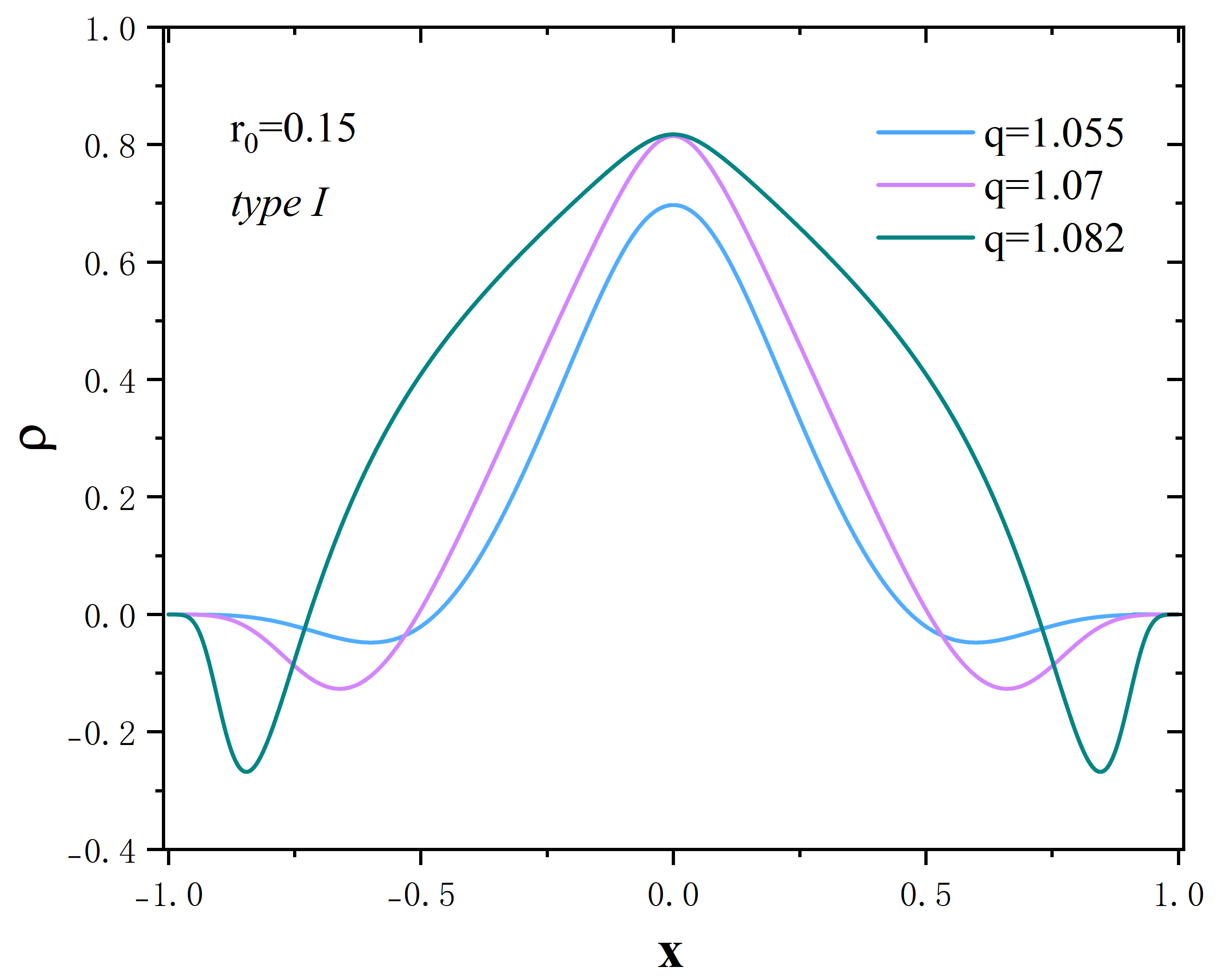}}
\subfigure[~]{\label{r2}
\includegraphics[width=0.31\textwidth]{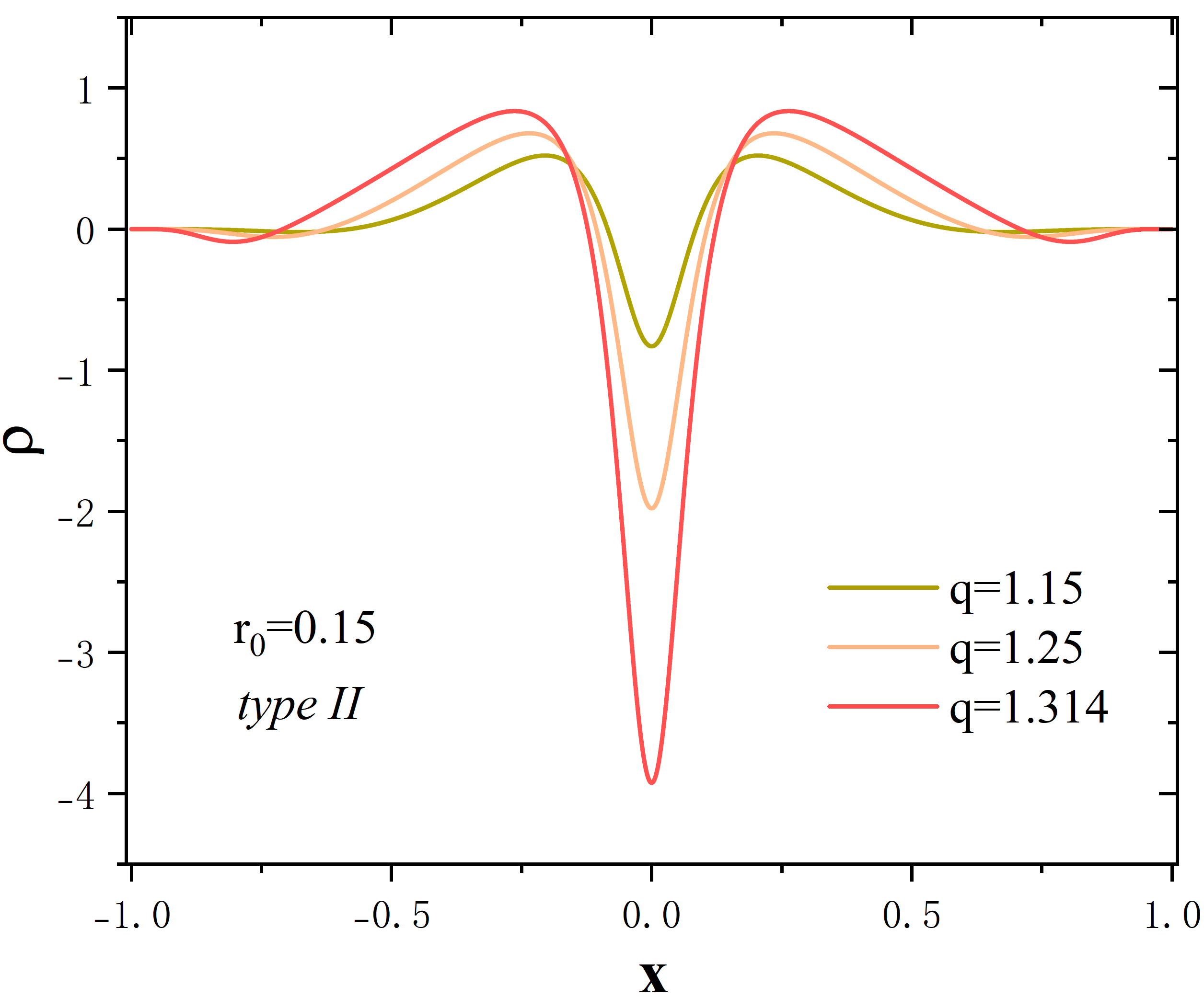}}
\subfigure[~]{\label{r3}
\includegraphics[width=0.32\textwidth]{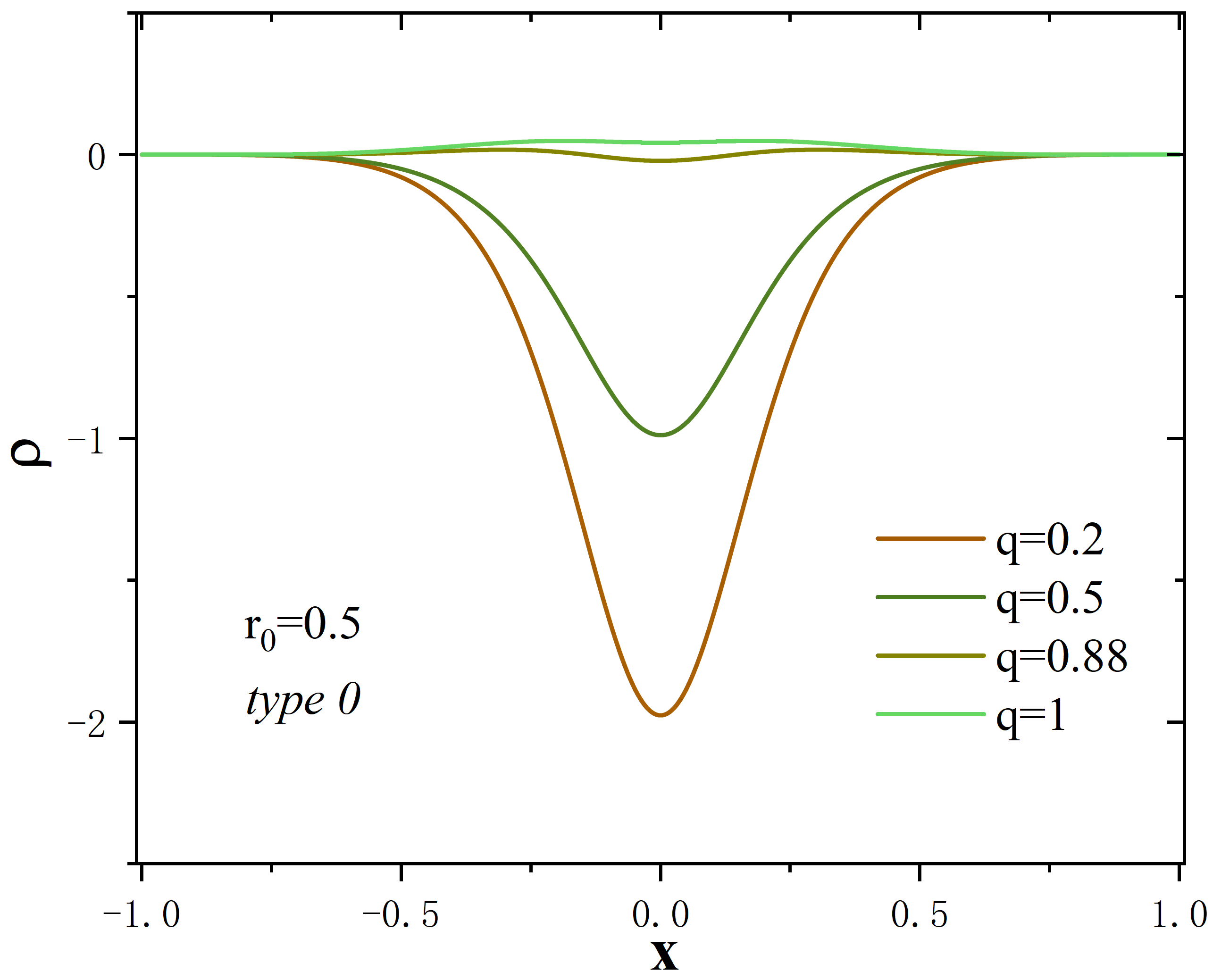}}
  \end{center}
\caption{ The energy density $\rho$ versus coordinate $x$ for different $q$ . Fig. (a) and (b) correspond to the negative mass regions of \textit{type I} and \textit{type II} solutions when $r_{0}$=0.15. Fig. (c) represents the positive mass solution \textit{type 0} for $r_{0}$=0.5. }
\label{rou}
\end{figure}

Traversable wormholes with phantom scalar fields violate the null energy condition. To satisfy the null energy condition it is necessary that $\rho + p_{i}\ge 0$. It can be seen from Fig. \ref{nec} that $\rho + p_{1}< 0$ and $\rho + p_{2}> 0$ for both positive and negative mass solutions. In addition, we focus on the total energy density in the solution in Fig. \ref{rou}. For the negative mass regions of \textit{type I} and \textit{type II} solutions with $r_{0}=0.15$, the distribution of $\rho$ is positive in some areas and negative in other areas as shown in Fig. \ref{r1} and Fig. \ref{r2}. For the positive mass solution \textit{type 0} corresponding to $r_{0}=0.5$, when $q$ is large, the energy density is positive, and as $q$ decreases, the contribution of the nonlinear electromagnetic field decreases causing $\rho$ to become negative. Eventually, for small $q$, the contribution of the phantom field exceeds the magnetic monopole, and $\rho$ becomes completely negative. Similar conclusions hold for other $r_{0}$ values.

\subsection{Null geodesic equations and effective potential}

\begin{figure}
  \begin{center}
\subfigure[~]{\label{v1}
\includegraphics[width=0.33\textwidth]{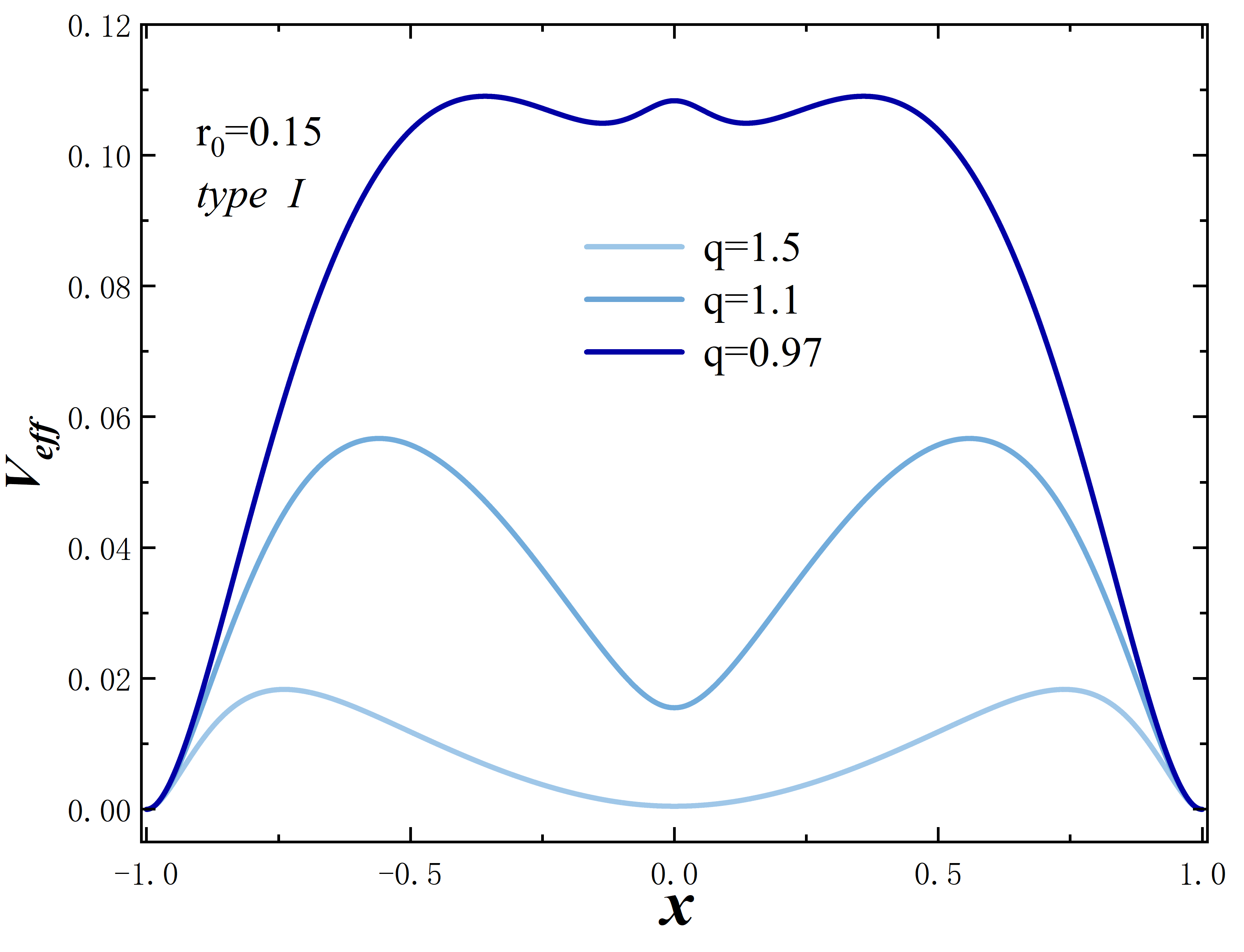}}
\subfigure[~]{\label{v2}
\includegraphics[width=0.33\textwidth]{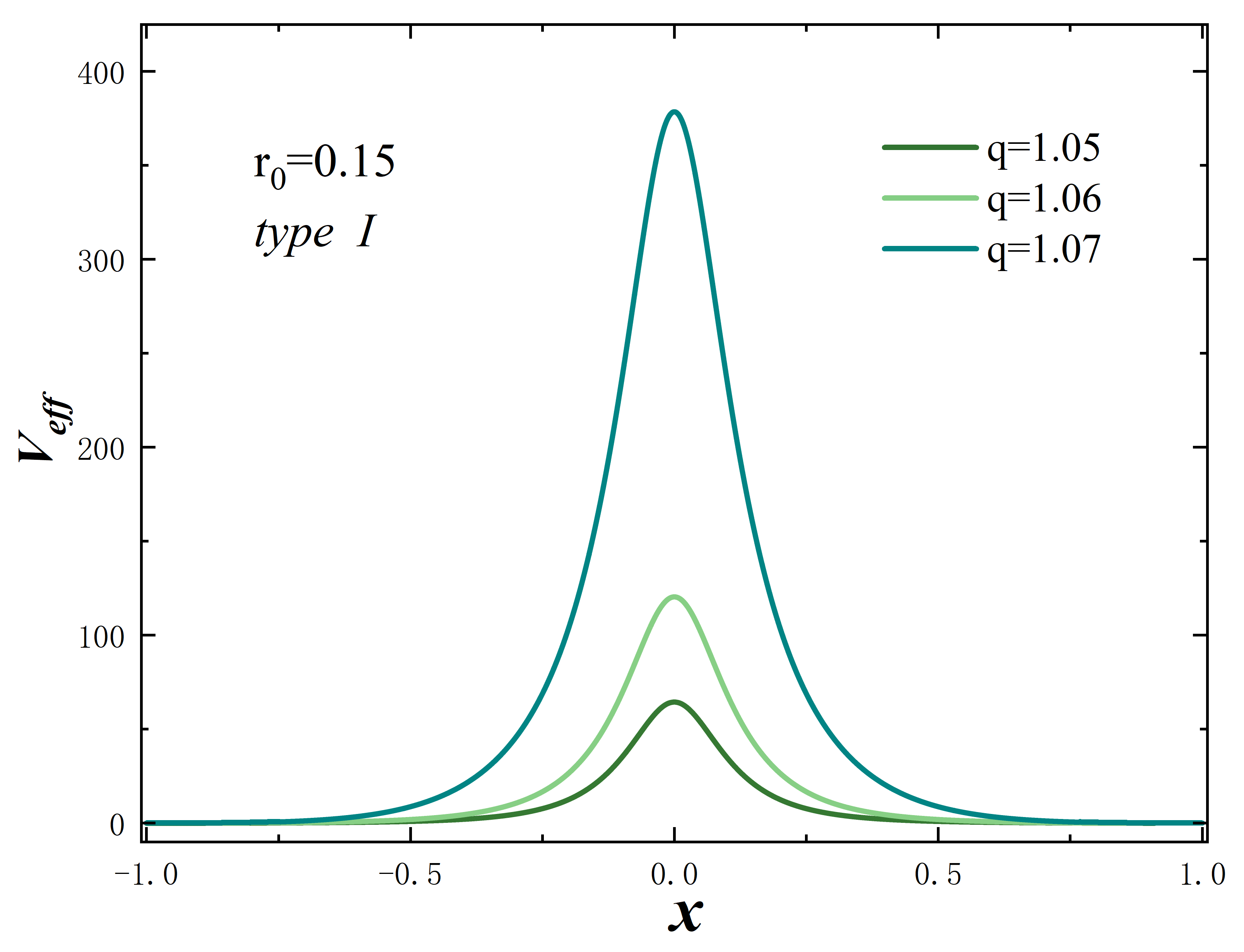}}
\subfigure[~]{\label{v3}
\includegraphics[width=0.32\textwidth]{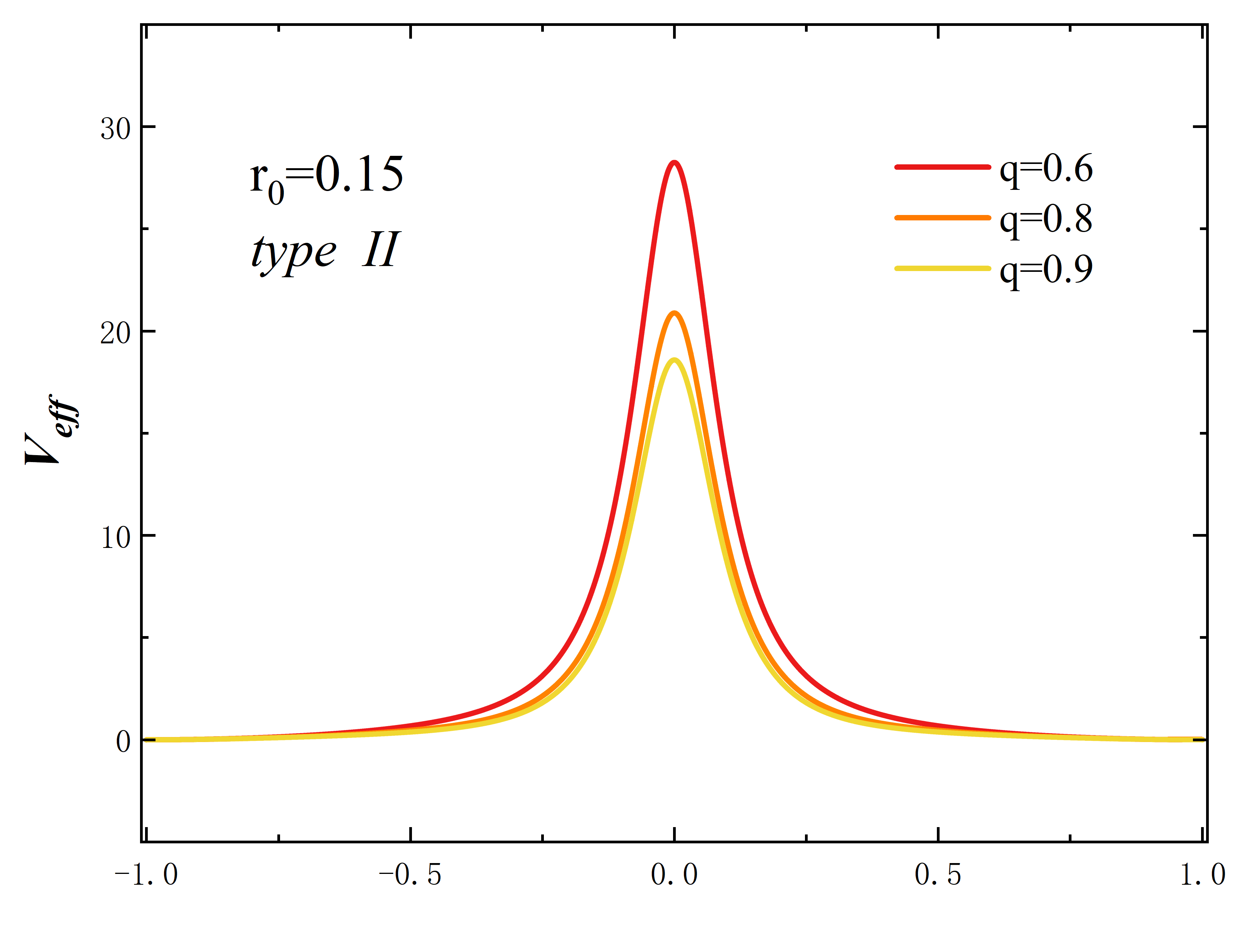}}
\subfigure[~]{\label{v4}
\includegraphics[width=0.32\textwidth]{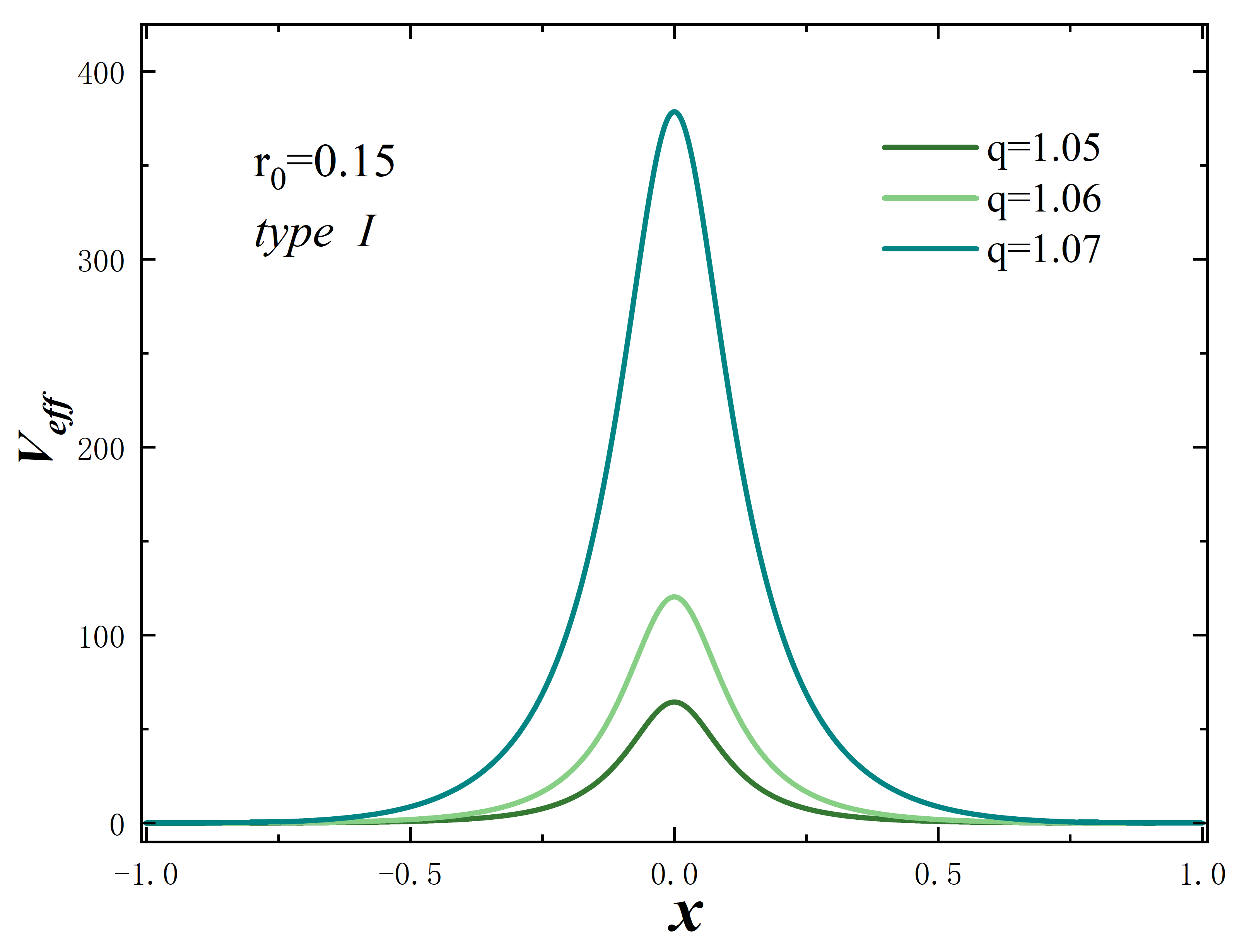}}
\subfigure[~]{\label{v5}
\includegraphics[width=0.32\textwidth]{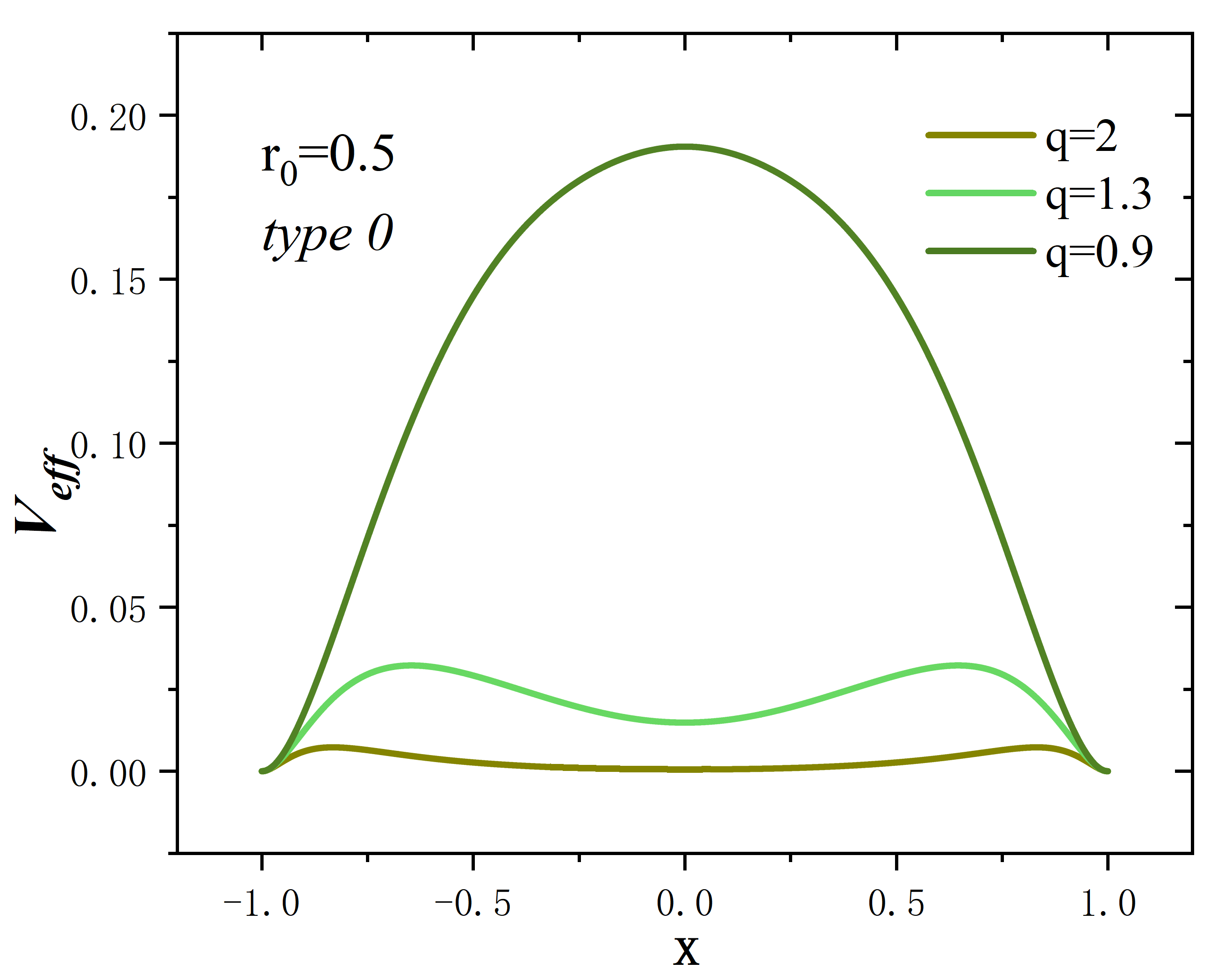}}
  \end{center}
\caption{ The effective potential $V_{eff}$ versus coordinate $x$ for different $q$. Fig. (a) and (b) correspond to the positive mass region and negative mass region of the \textit{type I} solution when $r_{0}$=0.15. Fig. (c) and (d) correspond to the positive mass region and negative mass region of the \textit{type II} solution when $r_{0}$=0.15. Figure (e) represents the positive mass \textit{type 0} solution for $r_{0}$=0.5. }
\label{veff}
\end{figure}

Photons in a gravitational field move along null geodesics which is given by
\begin{equation} 
g_{\mu \nu} \dot{x}^{\mu} \dot{x}^{\nu}=0,
\end{equation}
where in our case $x^{\mu}=\left ( t,r,\theta ,\varphi  \right ) $ and dots represent derivatives with respect to the affine parameter $\lambda$ along the geodesics. Due to the static character and the spherical symmetry of the system, we can assume that the orbit lies in the equatorial plane $\theta = \pi/2$ and defines two conserved quantities: $E=-g_{tt} \dot{t}$ and the angular momentum $E=g_{\varphi\varphi} \dot{\varphi}$. Consequently, taking into account the relevant metric components, the geodesic equation can be written as (re-scaling the affine parameter as $\lambda=\lambda/L$)

\begin{equation}
\dot{r}^{2}+\frac{1}{g_{tt}} \frac{1}{g_{rr}}(\frac{1}{b^{2}}+\frac{g_{tt}}{g_{\varphi \varphi }})=\dot{r}^{2}  +\frac{1}{C} (\frac{e^{B}}{Ce^{-B}h}-\frac{1}{b^{2}})=0,
\end{equation}
where we have defined the impact parameter as $b\equiv \frac{L}{E}  $  for a certain null particle. We define the effective potential $V_{eff}$ is

\begin{equation}
V_{eff}=\frac{e^{B}}{Ce^{-B}h}.
\end{equation}
When $r_{0}=0.15$, $V_{eff}$ of \textit{type I} and \textit{type II} solutions are plotted in Fig. \ref{v1} and Fig. \ref{v2}. For \textit{type I} solution, when $q$ takes a larger value, such as $q=1.5$, $V_{eff}$ has two maximum points and one minimum point. Then, as the value of $q$ decreases, the effective potential appears three maximum points and eventually becomes a maximum point as shown in Fig. \ref{v1}. When $q$ is incorporated into the negative mass solution, $V_{eff}$ still maintains a maximum point as shown in Fig. \ref{v2}. For \textit{type II} solution, $V_{eff}$ always has one maximum point which is plotted in Fig. \ref{v3} and Fig. \ref{v4}. $V_{eff}$ of the positive mass solution corresponding to $r_{0}=0.5$ is shown in Fig. \ref{v3}. It is shown that as $q$ decreases, $V_{eff}$ first has two maximum points and one minimum point, and finally becomes one maximum point. Where the extreme points $r_{LR}$ statisfy

\begin{equation}
\left.\frac{d V_{e f f}}{d r}\right|_{r_{L R}}=0.
\end{equation}
Therefore, for a null particle, if the square of the impact parameter $b_{c}^{2}=r_{L R}^{2} / n\left(r_{L R}\right) o^{2}\left(r_{L R}\right),$ photons coming from elsewhere will reduce the radial velocity to 0 at $r_{L R}$ and rotate around the wormhole. The corresponding orbits of the null particle are then referred to as the light ring. Moreover, for the maximum point of the effective potential, $V_{eff}(r_{LR})<0$, the light ring corresponding to $r_{LR}$ is unstable. On the contrary, the minimum point of the effective potential corresponds to a stable light ring.

\begin{figure}
  \begin{center}
\subfigure{\includegraphics[width=0.47\textwidth]{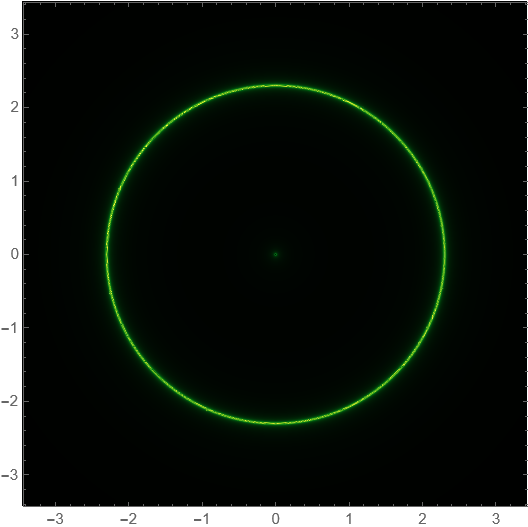}}
\subfigure{\includegraphics[width=0.48\textwidth]{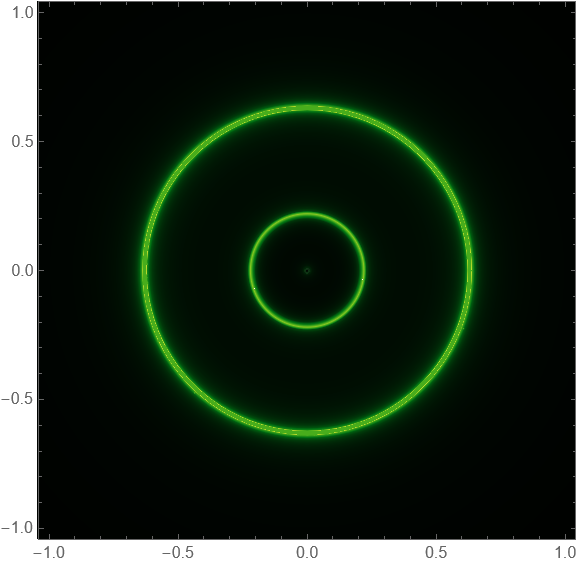}}
  \end{center}
\caption{The illustrations picture of the light ring corresponds to the \textit{type I} solution for $r_{0}=0.15$ with $q=1.5$, $M=1.4978$(left) and $q=0.97$, $M=0.6626$(right).}
\label{lr}
\end{figure}

Because our wormhole model is symmetric, the distribution of the light rings on both sides of the wormhole is identical. Therefore, we only show the light ring diagram of spacetime on one side of the wormhole in Fig. \ref{lr}, taking the \textit{type I} solution $r=0.15$ as an example. It can be seen that there is always a light ring at the throat of the wormhole. In the following, we discuss the cases where the corresponding ADM mass is positive and negative respectively.

(i) ADM mass is positive: When $q$ is large, except for the light ring at the throat, there is one light ring. As $q$ decreases, two light rings appear. The images of light rings seen from the other side of the wormhole are identical.

(ii) ADM mass is negative: When $q$ corresponds to the negative mass region, there is no light ring, only a special unstable light ring at the throat. Because negative mass will bring about a repulsive effect, the repulsive effect of spacetime on both sides of the wormhole will form a balance at the throat, leading to the formation of the unstable light ring, so its formation mechanism is different from the formation mechanism of the positive mass spacetime light ring.

\subsection{Wormhole geometry}
Here we investigate the geometric properties of the wormhole. We take the \textit{type I} and \textit{type II} solutions with negative mass when $r_{0}=0.1$ as an example to draw the images. The method is to fix the $t$ and $\theta$ to obtain a two-dimensional hypersurface, which allows us to better observe the shape of the wormhole and the number of throats. Then we can embed the two-dimensional hypersurface into a three-dimensional Euclidean space, The resulting embedding map visualizes the wormhole geometry. This technique allows us to better understand the topology and properties of wormhole solutions.  

The specific steps are as follows. Fix the time $t$ and $\theta$($\theta=\pi/2$) and then use the cylindrical coordinates $(\rho ,\varphi,z )$, the line element can be written as

\begin{equation}
\begin{split}
ds^{2}&=Be^{-A} dr^{2}+Be^{-A} hd \varphi^{2}
\\&=d \rho^{2}+d z^{2}+\rho^{2} d \varphi^{2} .
\end{split}
\end{equation}

By comparing the metric forms in spherical coordinates and cylindrical coordinates, we can obtain expressions for $\rho$ and $z$
\begin{equation}
\rho (r)=\sqrt{B(r)e^{-A(r)}h(r)} ,  \ z(r)=\pm\int \sqrt{B(r)e^{-A(r)}-\left ( \frac{d\rho}{dr}  \right )^{2} } dr.
\end{equation}

Here $\rho$ corresponds to the circumferential radius, which corresponds to the radius of a circle located in the equatorial plane and having a constant coordinate $r$. The function $\rho$ has extreme points, where the first derivative is zero. When the second derivative of the extreme point is greater than zero, we call the point a throat, which corresponds to a minimal surface. When the second derivative of the extreme point is less than zero, we call the point the equator, which corresponds to the maximum surface.

\begin{figure}
  \begin{center}
\subfigure[~]{\label{cd1}
\includegraphics[width=0.45\textwidth]{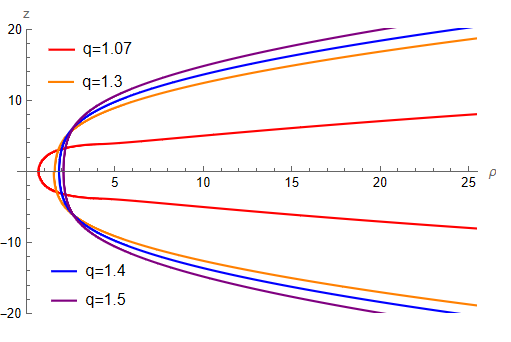}}
\subfigure[~]{\label{cd2}
\includegraphics[width=0.45\textwidth]{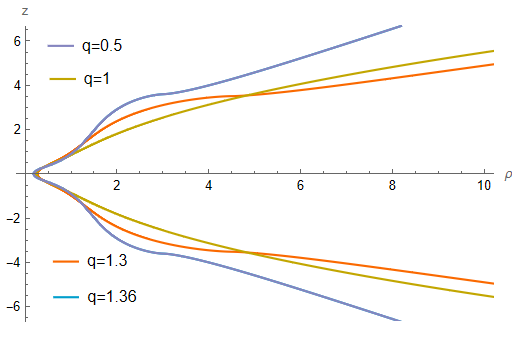}}
\subfigure[~]{\label{cd3}
\includegraphics[width=0.34\textwidth]{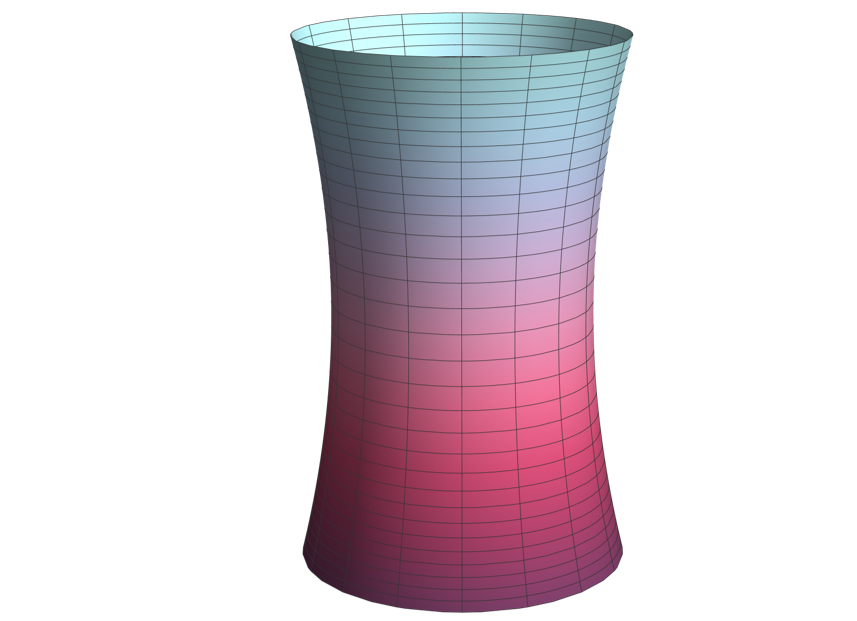}}
\subfigure[~]{\label{cd4}
\includegraphics[width=0.26\textwidth]{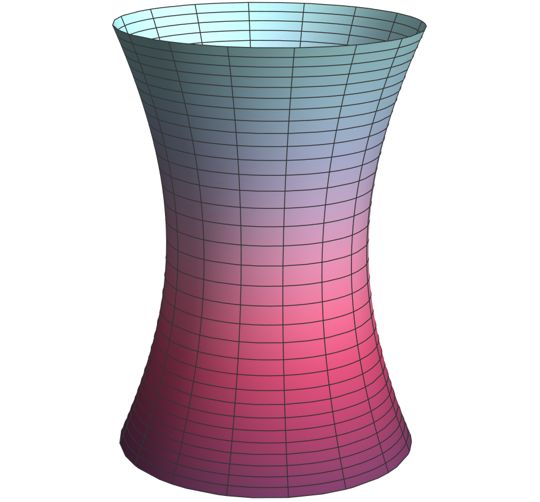}}
\subfigure[~]{\label{cd5}
\includegraphics[width=0.34\textwidth]{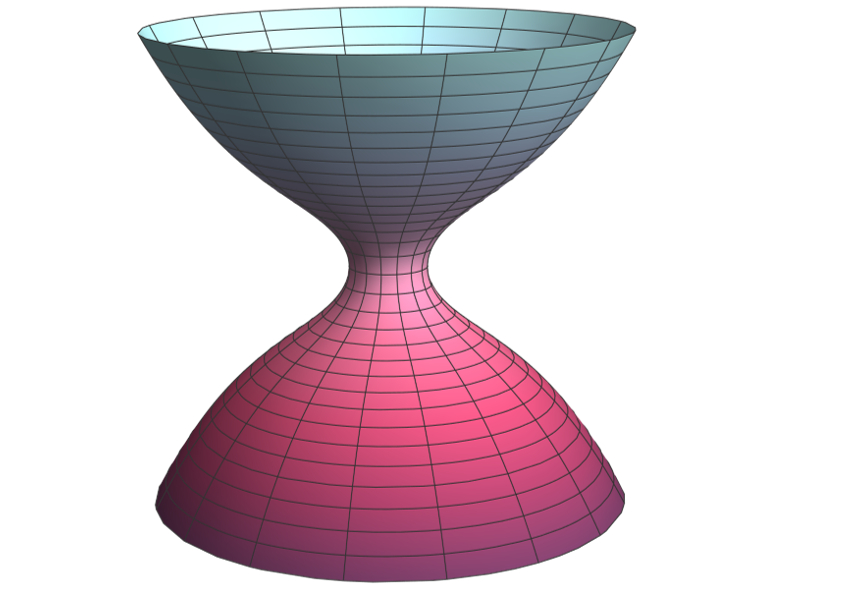}}
  \end{center}
\caption{ Wormhole geometry for $r_{0}=0.1$. (a) (b): The two-dimensional view of the isometric embedding of the equatorial plane of different $q$ for \textit{type I} and \textit{type II} solutions, respectively. (c): The three--dimensional plot with $q=1.07$, $M$= -0.466 for \textit{type I} solution. (d): The three--dimensional plot with $q=1.026$, $M$= 0.466 for \textit{type I} solution. (e): The three--dimensional plot with $q=1.36$, $M$= -11.012 for \textit{type II} solution. }
\label{wh}
\end{figure}

In Fig. \ref{wh}, we show two–dimensional views of the isometric embedding of the equatorial plane and the three–dimensional wormhole geometry for $r_{0}=0.1$. Fig. \ref{cd1} corresponds to the two-dimensional view of the \textit{type I} solution, and we plot the wormhole embedding diagram when negative mass appears in Fig. \ref{cd3} and positive mass wormhole is shown in Fig. \ref{cd4} for comparison. Fig. \ref{cd2} corresponds to the two-dimensional view of the \textit{type II} solution, and we plot the wormhole embedding diagram when negative mass appears in Fig. \ref{cd5}. It can be seen from the figure that no matter in the positive mass region or the negative mass region, the wormhole is always a single-throat symmetric wormhole.

\section{Conclusion}\label{sec5}
In this paper, we have investigated a wormhole model where Einstein’s gravity is coupled with nonlinear electromagnetic fields and the phantom field. Through the analysis of ADM mass, it is found that the model has one solution denoted as \textit{type 0} with constant positive ADM mass and two types of solutions with negative ADM mass denoted as \textit{type I} and \textit{type II}. Before this, whether the traversable wormhole model contains matter or not, negative energy density will be generated, leading to a series of violations of energy conditions, but no negative mass will be generated. The article \cite{Hao:2023kvf} couples the phantom field traversable wormhole with the nonlinear electromagnetic field of Bardeen spacetime and obtains negative ADM mass. Our model couples the wormhole with the nonlinear electromagnetic field of Hayward spacetime and also obtains negative ADM mass, which is consistent with the blue shift near the wormhole throat, showing that coupling nonlinear electromagnetic fields leads to negative mass.

In addition, we also study the effective potential and the light ring. Through the analysis of the effective potential, we find that for the \textit{type 0} solution, when $q$ is large, there is an unstable LR on each side of the wormhole and a stable LR at the throat. As $q$ decreases, there is only an unstable LR at the throat. For the \textit{type I} solution, when $q$ is large, there is an unstable LR on each side of the wormhole. As $q$ changes to the region where negative mass appears, the effective potential has only a maximum point at the throat, indicating that there is an unstable LR at $x=0$. For the \textit{type II} solution, no matter what value $q$ takes, the effective potential has only a maximum at the throat, indicating that there is always an unstable LR at the throat. It is worth noting that when spacetime presents negative mass, there is no stable LR and the unstable LR at the wormhole throat comes from the repulsive effect of spacetime on both sides of the wormhole, which is different from the unstable LR of positive mass spacetime.

Our work can be extended in some interesting ways in the future. On the one hand, our model can be extended to ADS spacetime, or explored in higher-order modified gravity, it is a worthwhile question whether negative mass will still appear in ADS spacetime, or in modified gravity theory. On the other hand, the dynamical
 stability analysis of the model is an important research topic, in particular, the emergence of negative mass may have some impact on the stability of the model. So in the future we can evolve the model to analyze its stability.

\section{ACKNOWLEDGEMENTS}

This work is supported by the National Key Research and Development Program of China (Grant No. 2020YFC2201503) and the National Natural Science Foundation of China (Grant No.12047501 and No.12275110).

\end{document}